\documentclass[12pt]{article}%
\usepackage{color}
\usepackage{fullpage}
\usepackage{amsmath}%
\usepackage{amssymb}%
\usepackage{amscd}%
\usepackage{amsthm}
\usepackage{graphicx}
\usepackage{amssymb}
\usepackage{epstopdf}
\usepackage{cancel}
\usepackage{setspace}
\usepackage{slashed}
\usepackage[hang,small]{caption}
\def\p{\partial}

\def\R{\mbb{R}}

\def\half{\frac{1}{2}}

\newcommand{\be}{\begin{equation}}
\newcommand{\ee}{\end{equation}}

\newcommand{\mbb}[1]{\mathbb{#1}}
\newcommand{\mrm}[1]{\mathrm{#1}}
\newcommand{\mcal}[1]{\mathcal{#1}}

\newcommand{\ph}[1]{\phantom{#1}}
\newcommand{\udten}[3]{#1^{#2}_{\ph{#2}#3}}
\newcommand{\duten}[3]{#1^{\ph{#2}#3}_{#2}}

\theoremstyle{definition}

\numberwithin{equation}{section}
\begin{document}
\begin{titlepage}
\bigskip
\rightline{}

\bigskip\bigskip\bigskip\bigskip
\centerline {\Large \bf {Holographic Confinement/Deconfinement Transitions}}
\centerline{\Large \bf{in Asymptotically Lifshitz Spacetimes}}
\bigskip\bigskip

\centerline{\large  Benson Way}
\bigskip\bigskip
\bigskip\bigskip
\centerline{\em Department of Physics, UCSB, Santa Barbara, CA 93106}
\centerline{\em benson@physics.ucsb.edu}
\bigskip\bigskip
\begin{abstract}Under a Scherk-Schwarz compactiÞcation, the AdS soliton and black brane provide a dual description to a confinement/deconfinement phase transition.  We extend this construction to asymptotically Lifshitz spacetimes.  In particular, we show that there must be a phase transition between the soliton and black hole if the solutions exist.  We also construct these solutions numerically and compute the phase diagram.    
\end{abstract}
\end{titlepage}


\onehalfspacing

\begin{section}{Introduction}
Since the emergence of the AdS/CFT correspondence \cite{Maldacena98,Gubser:1998bc,Witten:1998qj}, holography has become a powerful tool for understanding certain strongly coupled field theories.  In particular, there are many promising attempts to model interesting condensed matter systems with a gravity dual \cite{Hartnoll:2009sz,Herzog:2009xv,McGreevy:2009xe}.  

Most of these holographic models use a gravity dual which is asymptotically Anti-de Sitter (AdS).  The field theories described by these models are relativistic and have a conformal symmetry in the ultraviolet.  However, many condensed matter systems are non-relativistic.  For instance, quantum critical systems often exhibit a Lifshitz scaling symmetry:
\be\label{lifshitzscaling}
t\rightarrow\lambda^z t,\qquad x\rightarrow\lambda x\;,
\ee
where time and space scale anisotropically.  Such a scaling symmetry is modeled holographically by a Lifshitz metric \cite{Kachru:2008yh}:
\be\label{lifshitz1}
ds^2=\ell^2\left(-r^{2z}dt^2+\frac{dr^2}{r^2}+r^2 dx_idx^i\right)\;.
\ee
This metric exhibits the scaling \eqref{lifshitzscaling} if one also scales $r\rightarrow\lambda^{-1}r$.  
If $z=1$, this spacetime is AdS in Poincar\'e coordinates with AdS length scale $\ell$.  
\end{section}

Attempts were made to understand the holographic dictionary for asymptotically Lifshitz spacetimes \cite{Ross:2009ar,Ross:2011gu}.  To describe these systems at finite temperature, one must construct a black hole that asymptotically approaches \eqref{lifshitz1}.  Indeed, these types of solutions have been studied in \cite{Danielsson:2009gi,Mann:2009yx,Bertoldi:2009vn,Balasubramanian:2009rx,Bertoldi:2009dt,Amado:2011nd,Gonzalez:2011nz,Myung:2012xc}.

In this paper, we aim to study a new class of solutions that are asymptotically Lifshitz.  However, we require a new boundary condition.  We simply impose that one of the spatial directions be compactified to a circle (with fermions being antiperiodic around this circle, which manifestly breaks supersymmetry \cite{Scherk:1978ta}).  For $z=1$, there is a bulk solution with this boundary condition given by the AdS soliton \cite{Witten:1998zw,Horowitz:1998ha}.  In the case of asymptotically $AdS_5\times S^5$ solutions which are dual to $\mcal N=4$ super Yang-Mills, this boundary condition introduces masses for the fermions and scalars.  The AdS soliton is the gravitational dual to this state and represents the confining vacuum with a mass gap \cite{Witten:1998zw,Horowitz:1998ha,Surya:2001vj}.  This system also exhibits a phase transition at higher temperatures into a deconfining phase described by a black hole.  In the context of holographic superconductors  \cite{Gubser:2008px,Hartnoll:2008vx,Hartnoll:2008kx,Horowitz:2010gk}, the AdS soliton was used to model an insulator/superconductor transition \cite{Nishioka:2009zj,Horowitz:2010jq}.  

We would like to generalize the AdS soliton solution to $z>1$.  We opt to use the massive vector model considered in \cite{Ross:2009ar,Ross:2011gu} where the holographic dictionary has been developed.  A similar solution for $z=3$ was found in \cite{Gonzalez:2011nz} using a massive gravity theory in three bulk dimensions.  These solutions will also have the additional advantage that their zero temperature limit is well-defined.  If $z\neq1$, the Lifshitz spacetime suffers from a curvature singularity in the deep infrared in the form of diverging tidal forces \cite{Hartnoll:2009sz,Kachru:2008yh,Copsey:2010ya,Horowitz:2011gh}.  The zero temperature limit of black hole solutions will also suffer from this singularity.  While the true nature of this singularity has yet to be understood, it may pose a problem for understanding the deep IR physics of the theory.  In the soliton solutions, the spacetime caps off and there is no curvature singularity, even at zero temperature.  See \cite{Harrison:2012vy,Bao:2012yt} for other ways of resolving the singularity.  

In the following section, we present the massive vector model containing Lifshitz solutions and review the AdS soliton and its confinement/deconfinement transition in $z=1$.  Then in section 3, we  present an ansatz which we used to study the equations of motion.  In section 4, we discuss the thermodynamics of the system where we prove that a confinement/deconfinement transition exists for any $z$ where solutions also exist.  We then construct these solutions numerically and compute a phase diagram in section 5.  We finish with a few closing remarks.  

\begin{section}{Action and AdS Solutions}
For condensed matter applications, we would like to model a field theory in (2+1) dimensions.  Because of the extra holographic direction and the compactified circle, we choose to work in five bulk dimensions.  It is straightforward to generalize our analysis to any number of dimensions.  Therefore, consider the following action in five dimensions,
\be\label{action}
S=\int d^5x\;\sqrt{-g}\left(R-2\Lambda-\frac{1}{4}F_{\mu\nu}F^{\mu\nu}-\frac{m^2}{2}A_\mu A^\mu\right)\;,
\ee
where $F=dA$.  The equations of motion from this action are
\begin{align}\label{eom}
R_{\mu\nu}&=\frac{2\Lambda}{3}g_{\mu\nu}+\half F_{\mu\lambda}\duten{F}{\nu}{\lambda}+\frac{m^2}{2}A_\mu A_\nu-\frac{1}{12}F_{\rho\sigma}F^{\rho\sigma} g_{\mu\nu}\nonumber\\
\nabla_\mu F^{\mu\nu}&=m^2 A^\nu\;.
\end{align}
The action \eqref{action} is the five-dimensional version of the one considered in \cite{Ross:2009ar,Ross:2011gu}, where much of the holographic dictionary has been developed.  

One solution to \eqref{eom} is the Lifshitz spacetime
\be\label{lifshitz2}
ds^2=\frac{\ell^2}{\rho^2}\left(-\rho^{-2(z-1)}dt^2+d\rho^2+d\eta^2+dx^2+dy^2\right)\;.
\ee
This metric is equivalent to the one presented in $\eqref{lifshitz1}$ with $r=1/\rho$.  The other fields and constants are given by
\begin{align}\label{lifsol}
A&=\frac{\ell}{\rho^z}\sqrt{\frac{2(z-1)}{z}}dt\nonumber\\
\Lambda&=-\frac{z^2+2z+9}{2\ell^2}\nonumber\\
m^2&=\frac{3z}{\ell^2}\;.
\end{align}

Now let us set $z=1$ in \eqref{lifsol}.  Then the equations of motion reduce to those of pure Einstein gravity with a negative cosmological constant.  We therefore have the usual asymptotically AdS solutions.  In particular, we have the planar black hole, or black brane metric
\be\label{planar}
ds^2=\frac{\ell^2}{\rho^2}\left(-f(\rho)dt^2+\frac{d\rho^2}{f(\rho)}+d\eta^2+dx^2+dy^2\right),\qquad f(\rho)=1-\frac{\rho^4}{\rho_+^4}\;,
\ee
where $\rho_+$ is the horizon radius.  This black hole has temperature and free energy density\footnote{Here, we are using units where $16\pi G=1$.}
\be\label{tfbh}
T= \frac{1}{\pi \rho_+}\;,\qquad \mcal F=-\frac{\ell^3}{\rho_+^4}=-\pi^4\ell^3T^4\;.
\ee
We can also make the coordinate $\eta$ in \eqref{planar} periodic with any period we wish.  If we then perform a double Wick rotation $(dt\rightarrow id\eta,d\eta\rightarrow idt)$, we obtain the AdS soliton \cite{Witten:1998zw,Horowitz:1998ha}
\be\label{adssoliton}
ds^2=\frac{\ell^2}{\rho^2}\left(-dt^2+\frac{d\rho^2}{f(\rho)}+f(\rho)d\eta^2+dx^2+dy^2\right)\;,\qquad f(\rho)=1-\frac{\rho^4}{\rho_0^4}\;.
\ee
In order to avoid a conical singularity at $\rho=\rho_0$, the coordinate $\eta$ in \eqref{adssoliton} must have a period
\be
\gamma=\pi\rho_0\;.
\ee
The AdS soliton takes the shape of a cigar with a tip at $\rho=\rho_0$.  Since the spacetime only exists for $\rho\leq\rho_0$, the dual field theory is confining and has a mass gap \cite{Witten:1998zw}.  We can consider the AdS soliton at any temperature with a free energy density given by
\be\label{fsoliton}
\mcal F=-\frac{\ell^3}{\rho_0^4}=-\frac{\pi^4\ell^3}{\gamma^4}\;.
\ee

Together, the black hole (with the coordinate $\eta$ having period $\gamma$) and the soliton are geometries with the same asymptotics ($\R^{1,2}\times S^1)$.  At finite temperature, they form two competing phases.  These solutions have the same Euclidean geometries, differing only in the interpretation of which $S^1$ is Euclidean time and which is a spatial direction.  To find the critical temperature, we compare the free energies \eqref{tfbh} and \eqref{fsoliton}.  Since the free energies are computed from the Euclidean action, we find, unsurprisingly, a phase transition when $T=1/\gamma$ with the soliton phase being preferred at lower temperature.  This is a first-order phase transition, analogous to the Hawking-Page transition\footnote{An analogous transition between asymptotically Lifshitz black holes and thermal Lifshitz was considered in \cite{Amado:2011nd}} \cite{Hawking:1982dh}.  In the dual field theory, this is a confinement/deconfinement transition.  

We would now like to extend these spacetimes and the corresponding phase transition to asymptotically Lifshitz solutions with $z>1$.  Unlike the $z=1$ case (pure AdS), the pure Lifshitz metric \eqref{lifshitz2} is not invariant under a double Wick rotation.  In parciular, the analytic continuation of a black hole solution would give a spacetime with different asymptotics.  Therefore, the black hole solutions and solitons must be constructed independently\footnote{A double Wick rotation of a Lifshitz black hole would yield an anisotropic soliton.  Similarly, a double Wick rotation of a Lifshitz soliton will yield an anisotropic black hole.  While these solutions may be of interest, such a transformation would make the vector field in our model complex, so we do not discuss them here.}.  
\end{section}

\begin{section}{Ansatz and Boundary Conditions}
Black hole solutions in this model have been constructed before in \cite{Danielsson:2009gi,Mann:2009yx,Bertoldi:2009vn}, but we include the analysis here for completeness.  To find asymptotically Lifshitz black hole and soliton solutions, consider the following ansatz:
\begin{align}\label{ansatz}
ds^2&=\frac{\ell^2}{\rho^2}\left(-\rho^{-2(z-1)}F(\rho)dt^2+\frac{d\rho^2}{R(\rho)}+H(\rho)d\eta^2+dx^2+dy^2\right)\nonumber\\
A&=\frac{\ell}{\rho^z}\sqrt{\frac{2(z-1)}{z}}\phi(\rho)dt\;,
\end{align}
with $\Lambda$ and $m^2$ as given in \eqref{lifsol}.  For now, we are only concerned with analytic expressions.  For numerics, we will need to modify this ansatz, but we defer the discussion of performing numerics to section 5.

The equations of motion can be reduced to
\begin{align}\label{eom2}
F''&+\half\left[-\frac{F'}{F}+\frac{R'}{R}-\frac{H'}{H}-\frac{4(z-1)}{\rho}\right]F'+\frac{z+2}{\rho}\left[-\frac{R'}{R}+\frac{H'}{H}+\frac{2z(z-1)}{\rho(z+2)}\right]F\nonumber\\
&\qquad\qquad\qquad\qquad\qquad\qquad\qquad\qquad\qquad\qquad\qquad\qquad-\frac{2(z-1)(\rho\phi'-z\phi)^2}{z\rho^2}=0\nonumber\\
H''&+\half\left[-\frac{F'}{F}+\frac{R'}{R}-\frac{H'}{H}+\frac{2(z-1)}{\rho}\right]H'+\frac{3}{\rho}\left[\frac{F'}{F}-\frac{R'}{R}+\frac{2(z-1)\phi^2}{\rho RF}-\frac{2(z-1)}{\rho}\right]H=0\nonumber\\
\phi''&+\half\left[-\frac{F'}{F}+\frac{R'}{R}+\frac{H'}{H}-\frac{2(z+2)}{\rho}\right]\phi'+\frac{z}{2\rho}\left[\frac{F'}{F}-\frac{H'}{H}-\frac{R'}{R}+\frac{6}{\rho}\left(1-\frac{1}{R}\right)\right]\phi=0\nonumber\\
R&=\frac{2zH\big[(z^2+2z+9)F+3(z-1)\phi^2\big]}{z\rho(\rho H'-6H)F'+2\big[-z(z+2)\rho FH'+(6z(z+1)F+(z-1)(\rho\phi'-z\phi)^2)H\big]}\;.
\end{align}
Note that the function $R$ can be expressed in terms of the other functions.  To preserve planar symmetry in the black holes, we set $H=1$.  There are thus two second-order ODEs in $F$ and $\phi$ for the black hole and three second-order ODEs in $F$, $H$, and $\phi$ for the soliton.  

Our ansatz \eqref{ansatz} gives a one dimensional reduced Lagrangian that generates the equations of motion \eqref{eom2}.  As in \cite{Bertoldi:2009vn,Bertoldi:2009dt}, we can study the symmetries of this Lagrangian to find the following conserved quantity:
\be
\p\left[\frac{z\rho FH'+3\big[-z\rho F'+2(z-1)\rho\phi\phi'+2(z-1)z(F-\phi^2)\big]H}{3z\rho^{z+3}}\sqrt{\frac{R}{HF}}\right]=0\;,
\ee
and so
\be\label{c0}
\frac{z\rho FH'+3\big[-z\rho F'+2(z-1)\rho\phi\phi'+2(z-1)z(F-\phi^2)\big]H}{3z\rho^{z+3}}\sqrt{\frac{R}{HF}}=C_0,
\ee
for some constant $C_0$.  Of course, one can also verify that this quantity is conserved via the equations of motion \eqref{eom2}.  This conserved quantity will be a key component in our derivation of a Smarr relation in section 4.  

Let us now discuss the boundary conditions, beginning with the boundary conditions at $\rho=\rho_+$ or $\rho=\rho_0$, corresponding respectively to the horizon and the of the tip of the soliton.   A black hole solution requires that $F$ (and $R$) vanish linearly as $\rho\rightarrow\rho_+$.  Regularity also requires that $\phi$ vanish linearly.  A soliton requires that $H$ (and R) vanish linearly as $\rho\rightarrow\rho_0$.  Regularity then imposes conditions on $F'(\rho_0)$ and $\phi'(\rho_0)$.  

Now we turn to boundary conditions near $\rho\rightarrow0$.  We demand that the functions approach the Lifshitz metric \eqref{lifshitz2} as $\rho\rightarrow0$.  Then for small $\rho$, any deviation from Lifshitz is perturbative.  Let us now linearize the equations about the Lifshitz solution
\begin{align}\label{linear}
F(\rho)&=1+\epsilon F_1(\rho)\nonumber\\
H(\rho)&=1+\epsilon H_1(\rho)\nonumber\\
\phi(\rho)&=1+\epsilon \phi_1(\rho)\;,
\end{align}
where $H_1=0$ for the black hole.  From the linearized equations of motion, we find that the functions have the following behavior near $\rho\rightarrow0$:
\begin{align}\label{falloffs}
F_1&\sim a\rho^{z+3}+a_-\rho^{\lambda_-/2}+a_+\rho^{\lambda_+/2}\nonumber\\
H_1&\sim b\rho^{z+3}\nonumber\\
\phi_1&\sim c\rho^{z+3}+c_-\rho^{\lambda_-/2}+c_+\rho^{\lambda_+/2}\;,
\end{align}
where $\lambda_\pm=z+3\pm\sqrt{9z^2-26z+33}$.  See figure \ref{fig1} for a plot of these powers as a function of $z$. The $c$ coefficients are related to the others via
\be
c=\frac{3a(2z^2+z+9)-b z(z+2)(z+3)}{6(z-3)(z-1)}\;,\qquad c_\pm=-\frac{z a_\pm\lambda_{\pm}}{4(z-3)(z-1)}\;.
\ee
Following the analysis in \cite{Ross:2009ar}, we also require that the $\rho^{\lambda_-/2}$ terms vanish.  For $z\geq3$, these terms diverge\footnote{The $z=3$ case introduces logarithms.} as $\rho\rightarrow0$;  for $1\leq z<3$, these terms can be interpreted as boundary data for the vector field \cite{Ross:2009ar}.  As we will see in the next section, the terms proportional to $\rho^{z+3}$ are related to the energy density.  
\begin{figure}[t]
\centerline{\includegraphics[width=.45\textwidth]{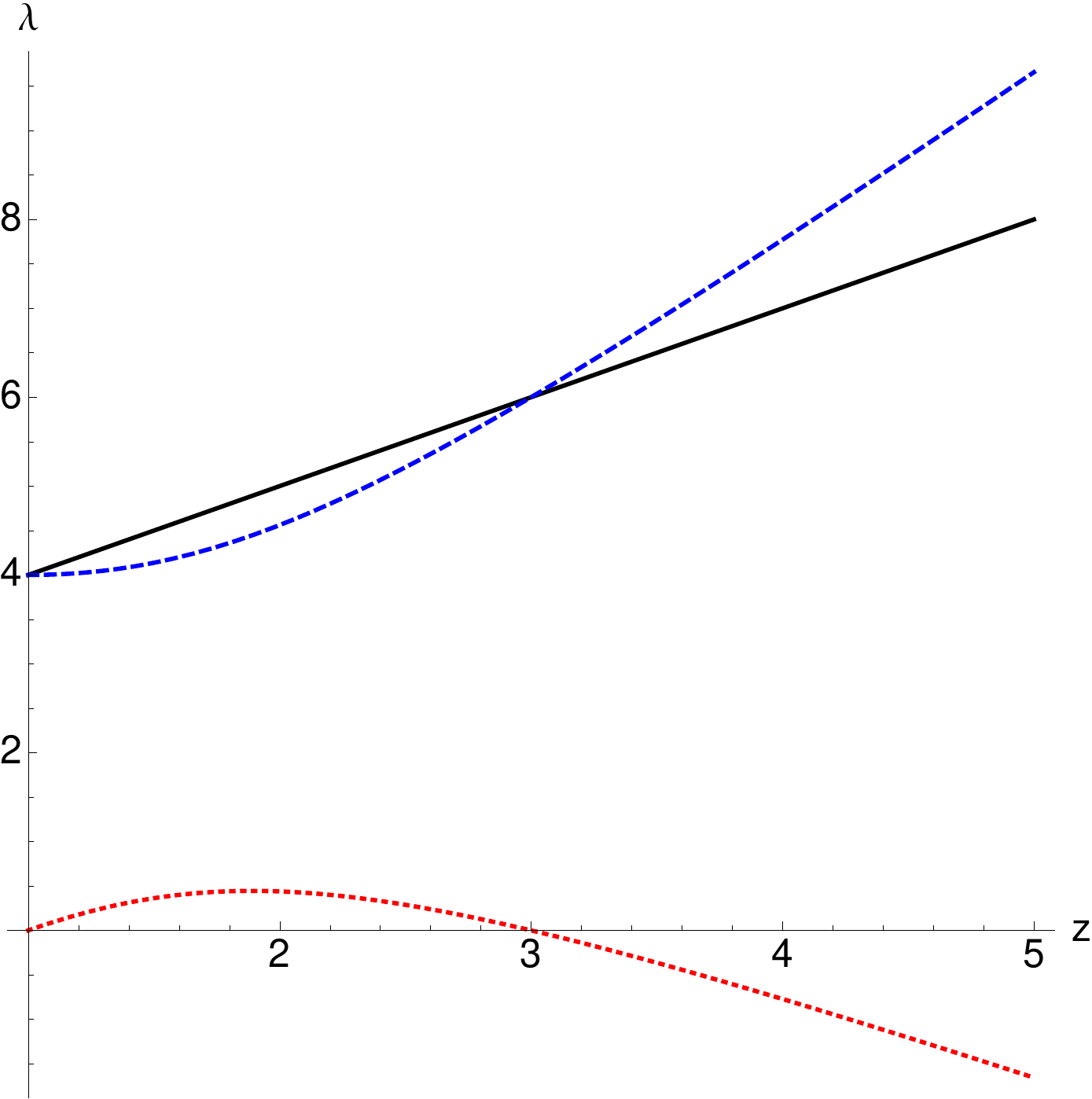}}
\caption{A plot of the various powers $\rho^\lambda$ appear in \eqref{falloffs}.  The dotted red line is $\lambda_-/2$, the blue dashed line is $\lambda_+/2$, and the solid black line is $z+3$.}
\label{fig1}
\end{figure}

For any given $z$, there are thus a family of black holes and solitons parametrized by $\rho_+$ and $\rho_0$, respectively.  These solutions are related to each other by a useful scaling symmetry:
\be\label{scaling}
t\rightarrow\lambda^{z}t\;,\qquad(\rho,x,y,\eta)\rightarrow \lambda(\rho,x,y,\eta)\;.
\ee  
Given a black hole solution with horizon at $\rho=\rho_+$, the above transformation will give another solution with horizon at $\lambda^{-1}\rho_+$.  Similarly, a soliton with tip at $\rho=\rho_0$ can be scaled to give a soliton with tip at $\rho=\lambda^{-1}\rho_0$.  
\end{section}
\begin{section}{Thermodynamics}
In this section, we discuss the thermodynamics of the black holes and solitons.  In particular, we will derive the thermodynamic Smarr relation found in \cite{Bertoldi:2009dt} and a similar relation for the soliton.  This will allow us to prove the existence of a phase transition, so long as both solutions exist.  For simplicity, we will henceforth set $16\pi G=\ell=1$.

We begin by computing the energy density.  The expression for the energy density in asymptotically Lifshitz spacetimes has been worked out in four dimensions in \cite{Ross:2009ar} from a non-relativistic boundary stress tensor complex.  Extending their work to five dimensions, the energy density is given by the $\rho\rightarrow0$ limit of the expression:
\be\label{energy}
\mcal E=2 \udten{s}{t}{t}-s^t A_t=-2\rho^{2z}s_{tt}+\rho^z\sqrt{\frac{2(z-1)}{z}} s_t\;,
\ee
where 
\begin{align}
s_{\mu\nu}&=\sqrt{-h}\left[(\pi_{\mu\nu}+3h_{\mu\nu})+\frac{z}{2}\sqrt{\frac{2(z-1)}{z}}(-A_\lambda A^\lambda)^{-1/2}(A_\mu A_\nu-A_\sigma A^\sigma h_{\mu\nu})\right]\nonumber\\
s_\nu&=-\sqrt{-h}\left[n^\mu F_{\mu\nu}-z\sqrt{\frac{2(z-1)}{z}}(-A_\mu A^\mu)^{-1/2}A_\nu\right]\;.
\end{align}
Here, $h_{\mu\nu}$ is an induced metric at some constant $\rho$; $n^\mu$ is a unit vector normal to the boundary and directed outwards; and $\pi_{\mu\nu}=K_{\mu\nu}-K h_{\mu\nu}$, where $K_{\mu\nu}=\nabla_{(\mu}n_{\mu)}$ is the extrinsic curvature of the boundary.   

Given the behavior near the boundary \eqref{linear} and \eqref{falloffs}, the energy density is
\be\label{energy2}
\mcal E=\frac{3+z}{z}(-3a+zb)\;.
\ee
We would like to relate this quantity to other quantities at the horizon or tip.  As in \cite{Bertoldi:2009dt}, we do this by evaluating the conserved quantity $C_0$ given by \eqref{c0}.  At the boundary, we find that $C_0$ satisfies
\be\label{ec0}
\mcal E=\frac{3}{3+z} C_0\;.
\ee

Fo derive the Smarr relation for the black holes, we evaluate $C_0$ at the horizon and then relate it to \eqref{ec0}.  At the horizon,
\be\label{c0horizon}
\mcal C_0=\frac{1}{\rho_+^{z+2}}\sqrt{F'(\rho_+)R'(\rho_+)}=Ts\;,
\ee
where $T$ is the black hole temperature, and $s$ is the entropy density.  Together with \eqref{ec0}, this implies the Smarr relation
\be\label{ebh}
\mcal E_{bh}=\frac{3}{z+3}Ts\;,
\ee
which is the five dimensional version of the same thermodynamic relation derived previously in \cite{Bertoldi:2009dt}.  

We can repeat this procedure for the soliton.  At the tip,
\be\label{c0tip}
C_0=-\frac{1}{3\rho_0^{z+2}}\sqrt{F(\rho_0)R'(\rho_0)H'(\rho_0)}\;.
\ee
By analogy with \eqref{c0horizon}, we can rewrite \eqref{c0tip} as $C_0=-\tau/3\gamma$, where $\gamma$ is the period of the soliton, and $\tau$ is the tension.  If we were to double Wick rotate the soliton, $\tau$ would be the entropy density of the resulting anisotropic black hole, and $1/\gamma$ would be its temperature.  From this and \eqref{ec0}, we find that
\be\label{esl}
\mcal E_{sl}=-\frac{1}{z+3}\frac{\tau}{\gamma}\;.
\ee

We can compute the free energy density via
\be\label{fenergy}
\mcal F=\mcal E-Ts\;.
\ee
Since the soliton has vanishing entropy, we see immediately from \eqref{esl} that the free energy is negative, and so the soliton is always preferred over thermal Lifshitz, as expected.  The free energy of the black holes are also negative and given by $\mcal F=-zTs/(z+3)$.  Equating the two free energies, we see that there is a phase transition when
\be\label{tcrit}
Ts=\frac{\tau}{z\gamma}\;.
\ee
From the scaling relations \eqref{scaling}, one can show that decreasing the black hole temperature also decreases the entropy density, so there always exists a temperature where \eqref{tcrit} can be satisfied.  We therefore conclude that if soliton and black hole solutions exist for a given z, there must always be a phase transition between them.  
\end{section}
\begin{section}{Numerics and Results}
Now we construct the black holes and solitons numerically.  To improve numerics, we must make modifications to our ansatz \eqref{ansatz}.  We must satisfy all of the boundary conditions and keep the functions at least twice differentiable.  For the black holes, we set
\begin{align}
F(\rho)&=(1-\rho^{2z+2})(1+\rho^z f_F(\rho))\nonumber\\
H(\rho)&=1\nonumber\\
\phi(\rho)&=(1-\rho^{2z+2})(1+\rho^z f_\phi(\rho))
\end{align}
The horizon is located at $\rho=\rho_+=1$.  The scaling \eqref{scaling} lets us obtain solutions for other values of $\rho_+$.  Here, the equations of motion give boundary conditions on $f'_F(1)$ and $f_\phi'(1)$.  At the boundary, we require $f_F(0)=f_\phi(0)=0$.  This ensures that the $\lambda_-$ terms in \eqref{falloffs} vanish.  The power of $2z+2$ was chosen so that the other modes in \eqref{falloffs} are unaffected.  

For the solitons, we set
\begin{align}
F(\rho)&=1-\rho^z g_F(\rho)\nonumber\\
H(\rho)&=1-\rho^{z+2} g_H(\rho)\nonumber\\
\phi(\rho)&=1-\rho^z g_\phi(\rho)
\end{align}
If we set $g_H(1)=1$, the tip is located at $\rho=1$.  The equations of motion then give boundary conditions on $g_F'(1)$ and $g_\phi'(1)$.  As in the black holes, we require $g_F(0)=g_H(0)=g_\phi(0)=0$ at the boundary.

Based on the behavior near $\rho\rightarrow 0$ given by \eqref{falloffs}, all of our functions are guaranteed to be at least twice differentiable.  

We opt to use a Newton-Raphson relaxation procedure on a finite difference grid.  Because of the inherent non-analytic behavior of the functions, particularly at $\rho=0$, we use a second-order central differencing in the interior of the grid and forward and backward differencing on the edges, rather than some higher order method.  

As a test of convergence of our code, we use the quantities
\be\label{convergence}
\left| 1-I_N(f_F,f_\phi)/I_{N+1}(f_F,f_\phi)\right|,\qquad\left| 1-I_N(g_F,g_H,g_\phi)/I_{N+1}(g_F,g_H,g_\phi)\right|\;,
\ee
where $I_N(f_i)$ is the average of the integrals of the functions $f_i$, computed using a grid of $N$ points.  As we see in the left plot in figure \ref{fig2}, we find the expected quadratic convergence.  

We can also use the relations \eqref{ebh} and \eqref{esl} to test the accuracy of our numerics.  We compute the energy density from the boundary $\mcal E_\p$ using \eqref{energy2}.  We then compute the energy from quantities at the horizon or tip $\mcal E_0$ using \eqref{ebh} and \eqref{esl}.  A measure of our accuracy is then given by $\mrm{Max}(|1-\mcal E_\p/\mcal E_0|,|1-\mcal E_0/\mcal E_\p|)$.  The result of this error is plotted in figure \ref{fig2} on the right.  We find that these relations are satisfied to 0.1\% accuracy.  
\begin{figure}[t]
\centerline{\includegraphics[width=.45\textwidth]{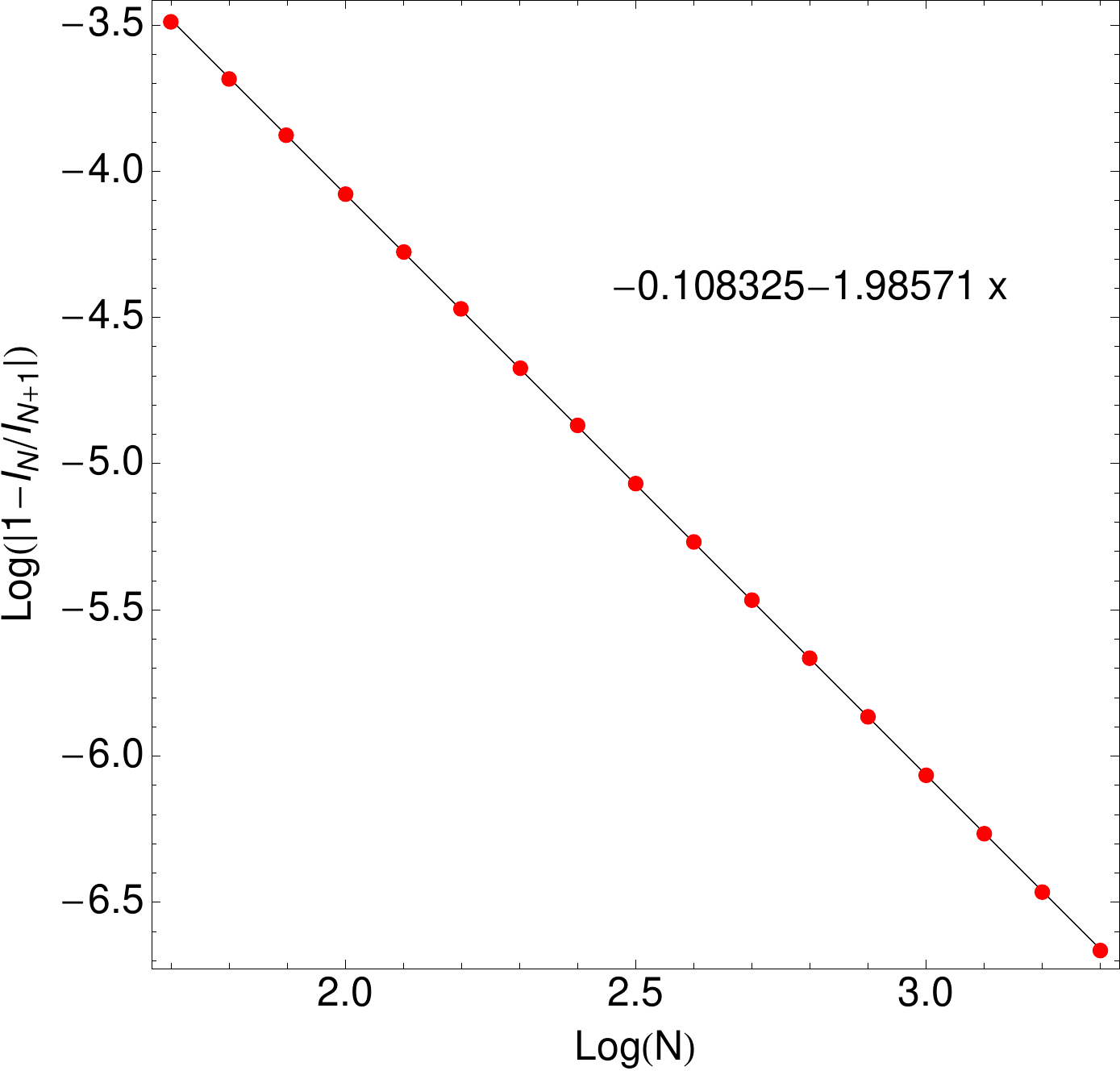}
\hspace{1cm}\includegraphics[width=.45\textwidth]{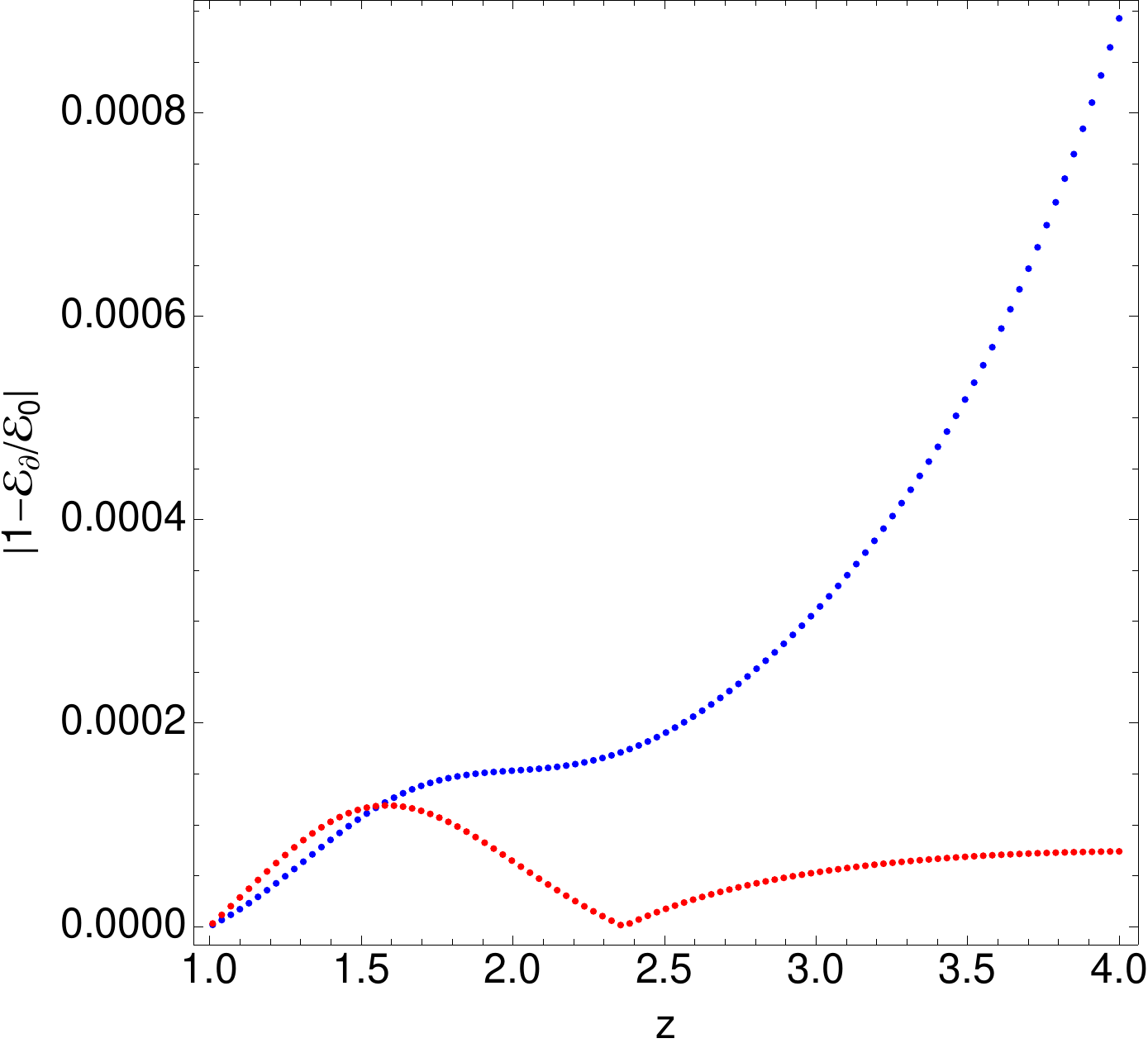}}
\caption{Left: test of quadratic convergence for the black hole by computing \eqref{convergence} at $z=1.3$.  The soliton is similar with a fit of $-0.107137-1.99508 x$. Right: the error in the Smarr relation \eqref{ebh} and the relation \eqref{esl} in red and blue, respectively.  Both plots are computed with $z=1.3$.}
\label{fig2}
\end{figure}

Now we compute the phase diagram using the following procedure.  First, we must fix a scale for the period of the spatial circle.  We use \eqref{scaling} to set the period of all of our soliton solutions to $\gamma=\pi$ and compute the free energy of these solutions.  For any given z, we then compute the free energy of the black hole and scale the solution so that it has the same free energy as the corresponding soliton.  Finally, we compute the temperature of the resulting black hole, giving us the critical temperature for the phase transition.  The result is plotted in figure \ref{fig3}.  We note that as $z\rightarrow1$, the critical temperature approaches the expected value of $1/\pi$.  
\begin{figure}[t]
\centerline{\includegraphics[width=.6\textwidth]{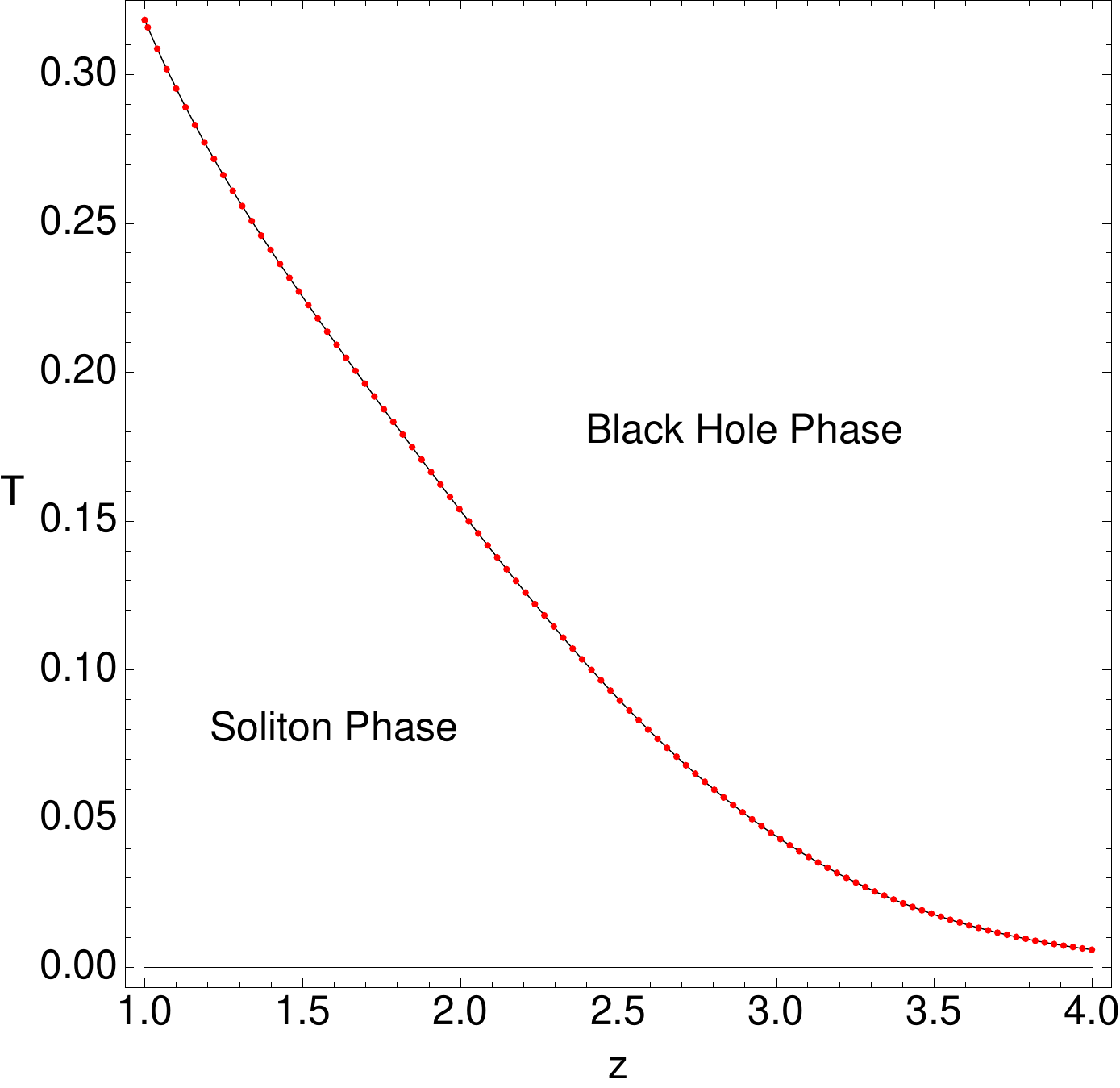}}
\caption{The phase diagram for the Lifshitz black hole/soliton system.  The period of the soliton has been scaled to $\gamma=\pi$, and we have set $16\pi G=\ell=1$.}
\label{fig3}
\end{figure}
\end{section}
\begin{section}{Discussion}
Using a massive vector model in five dimensions, we have shown that a confinement/deconfinement transition exists whenever black hole and soliton solutions exist.  It would be interesting to see if this analysis can be extended to other models that include Lifshitz metrics, especially models with embeddings in supergravity or string theory.  

We have also constructed Lifshitz solitons and black brane solutions numerically and computed their phase diagram.  It seems at first glance that the critical temperature is approaching zero as $z\rightarrow\infty$.  It is unclear whether the temperature continues to decrease for larger $z$, especially since numerics become more difficult (as can be seen from the right plot of figure \ref{fig2}).  Perhaps something can be said from the $z\rightarrow\infty$ limit of Lifshitz, $AdS_2\times \R^{D-2}$.  However, it is not clear how such a limit should translate to our black holes or solitons.  

One can consider holographic superconducting models of the Lifshitz solitons.  This may lead to a superconductor/insulator transition with $z>1$ Lifshitz scaling.  In the $z=1$ case, the AdS soliton represents an insulator, and can be made into a superconductor by varying a chemical potential \cite{Nishioka:2009zj}.  Together with the confinement/deconfinement transition, the system has a rich phase structure \cite{Horowitz:2010jq}.  One might suspect that these properties can be extended to $z>1$.  
\end{section}
\vskip 1cm
\centerline{\bf Acknowledgements}
\vskip .5 cm
It is a pleasure to thank Gary Horowitz and Jorge Santos for helpful discussions.  This work was supported in part by NSF grant PHY12-05500.

\singlespacing

\end{document}